# Optimizing Visual Cortex Parameterization with Error-Tolerant Teichmüller Map in Retinotopic Mapping

Yanshuai Tu[1][0000-0002-4619-2613], Duyan Ta[1],

Zhong-Lin Lu[2,3][0000-0002-7295-727X] and Yalin Wang[1*]

[1] Arizona State University, Tempe AZ 85201, USA
[2] New York University, New York, NY
[3] NYU Shanghai, Shanghai, China
[*] `ylwang@asu.edu`

**Abstract.** The mapping between the visual input on the retina to the cortical surface, i.e., retinotopic mapping, is an important topic in vision science and neuroscience. Human retinotopic mapping can be revealed by analyzing cortex functional magnetic resonance imaging (fMRI) signals when the subject is under specific visual stimuli. Conventional methods process, smooth, and analyze the retinotopic mapping based on the parametrization of the (partial) cortical surface. However, the retinotopic maps generated by this approach frequently contradict neuropsychology results. To address this problem, we propose an integrated approach that parameterizes the cortical surface, such that the parametric coordinates linearly relates the visual coordinate. The proposed method helps the smoothing of noisy retinotopic maps and obtains neurophysiological insights in human vision systems. One key element of the approach is the Error-Tolerant Teichmüller Map, which uniforms the angle distortion and maximizes the alignments to self-contradicting landmarks. We validated our overall approach with synthetic and real retinotopic mapping datasets. The experimental results show the proposed approach is superior in accuracy and compatibility. Although we focus on retinotopic mapping, the proposed framework is general and can be applied to process other human sensory maps.

## 1 Introduction

There is a great interest to understand, quantify, and simulate the human retinotopic mapping, i.e. the mapping between the visual field on the retina to the cortical surface. Since the first time functional magnetic resonance imaging (fMRI) was introduced to measure human retinotopic maps in vivo [1, 2], many improvements have been made: New experimental protocol were carried out, especially the traveling wave experiment [3]; The population receptive field model (pRF) was proposed to better interpret fMRI data [4]. Those researches bring inspiring insights to understand human vision systems.

Besides the great scientific interest, human retinotopic maps are also applied in several fields. In ophthalmology, there is a great need to evaluate the visual organization of amblyopia, which affects ~2% of all children [5] and may cause significant visual impairment if untreated. The retinotopic map is a better detection method and guides a proper treatment for amblyopia patients even for adults, which is usually believed incurable [6–8]. In neurology, fMRI signal and retinotopic maps have been adopted to help register cortical surfaces and discover more brain visual-related areas [9]. The retinotopic maps are also adopted in computer vision, e.g. a retinotopic Spiking Neural Network is inspired to recognize moving objects [10].

Although great progress has been made, there are several key characteristics, e.g. cortical magnification, have not been quantified precisely for human retinotopic maps due to several challenges. The fundamental challenge is the fMRI signal is of low signal-noise ratio and low spatial resolution, which makes the retinotopic maps noisy and blurring. Meanwhile, the cortical surface is complicated and convoluted, which makes the spatial smoothing extremely difficult for the noisy retinotopic maps.

Previous works tackle these challenges in separated steps [11–13]: the cortical surface is parametrized to a 2D domain, then smoothing methods are applied to the noisy retinotopic data, based on fitting smooth function on the parametrization domain. Although intuitive, the smoothed results by this approach are not compatible with the neuroscientific results: the retinotopic maps are topological (preserve the neighboring relationship) within each visual area in neuroscientific results [12, 14], yet the smoothed result does not preserve the neighboring relationship.

We address the parametrization and smoothing simultaneously and propose a novel cortical parametrization approach. The reason one shall combine them is that the parametrization influences the smoothing complexity, and vice versa. The smoothing is processed on the parametrized coordinates, where a small change of the coordinates may significantly reduce the complexity of smoothing in the 2D domain; On the other hand, high-quality visual coordinates can be used to guide the desired parametrization. Therefore, the combined approach has the potential to balance the difficulties of surface parametrization and smoothing. In specific, we first formulate the problem as a parameterization optimization problem. Then we solve the problem iteratively with Laplacian smoothing [15] and the proposed Error-Tolerant Teichmüller Map (ETTM). The ETTM can handle self-contradicting landmarks, so one can set landmarks for the parametrization with errors. This is the first approach that aligns the retinotopic parameters to the stimulus visual field and provides canonical space for retinotopic maps. The ETTM is proposed in general so it can be adapted to other sensory cortex like auditory maps [16].

## 2 Method

### 2.1 Background on Retinotopic Maps

We briefly explain the retinotopic mapping and introduce the notations. Suppose the visual stimulation at position $v = (v^{(1)}, v^{(2)})$ is $s(t, v)$. **Note** we take the polar angle system for the visual position, i.e. $v^{(1)}$ is the radical distance to the origin (i.e. eccen-tricity), and $v^{(2)}$ is the polar angle in the visual field (i.e. polar angle). The visual system will perception the stimulation and eventually activate a population of neurons, illustrated in **Fig. 1**. The main purpose of retinotopic mapping is to find the center $v$ and the extent $\sigma \in R^+$ of its receptive field for each point $P = (X, Y, Z) \in \mathbb{R}^3$ on the visual cortex. fMRI provides a noninvasive way to determine $v$ and $\sigma$ for $P$, based on the following procedure: (1) Design a stimulus $s(t; v)$, such that it is unique respect to location, i.e., $s(t; v_1) \neq s(t; v_2), \forall v_1 \neq v_2$;

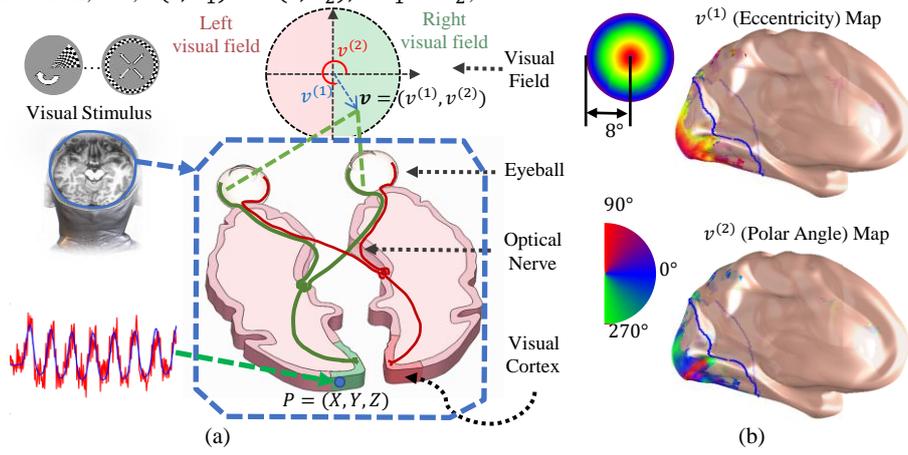

**Fig. 1.** (a) Illustration of the human visual system and retinotopic mapping procedure: The Visual Stimulus is presented in front of the subject's visual field, and then recording the fMRI signal during the process; (b) The retinotopic coordinates $v = (v^{(1)}, v^{(2)})$ are decoded, and rendered on the inflated cortical surface by population receptive field (pRF) analysis: The top and bottom are the visual eccentricities $v^{(1)}$ and visual polar angles $v^{(2)}$ on the inflated cortical surface, respectively.

(2) Present the stimulus sequence to an individual and record the fMRI signals during the stimulation; (3) For each point $P$ on cortical, collect the fMRI signal along the time, $y(t; P)$; (4) Determine the parameters, including its central location $v$ and its size $\sigma$, that most-likely generated the fMRI signals. Specifically, one assumes the neurons' spatial response $r(v'; v, \sigma)$ model, and the hemodynamic function $h(t)$, to predict the fMRI signal by, $\hat{y}(v, \sigma) = \beta(\int r(v'; v, \sigma)s(t, v')dv') * h(t)$, where $\beta$ is a coefficient that converts the units of response to the unit of fMRI activation. The parameters $v$ and $\sigma$ are estimated by minimizing the prediction error, i.e. $(v, \sigma) = \arg\min|\hat{y}(v, \sigma) - y(t; P)|^2$. The retinotopic maps are obtained when $(v, \sigma)$ is solved for every point on the cortical surface. **Fig. 1**(b) shows a typical retinotopic mapping decoded and rendered on the inflated cortical surface. Besides the retinotopic maps, one can further evaluate the goodness of retinotopic parameters for each location by computing the variance explained, $R^2 = \int |\hat{y} - \bar{y}|^2 dt / \int |y - \bar{y}|^2 dt$.

## 2.2 Problem Statement

We wish to parametrize the visual cortex such that the parametric coordinates, $\hat{u} = (\hat{u}^{(1)}, \hat{u}^{(2)})$ in polar coordinates linearly relate the retinotopic coordinates $v = (v^{(1)}, v^{(2)})$. Namely, $\hat{\rho} = k_1 v^{(1)} + b_1$, $\hat{\theta} = k_2 v^{(2)} + b_2$, where $\hat{\rho} = |\hat{u}^{(1)} + i\hat{u}^{(2)}|$, $\hat{\theta} = \arg(\hat{u}^{(1)} + i\hat{u}^{(2)})$, $k = (k^{(1)}, k^{(2)})$, and $b = (b^{(1)}, b^{(2)})$ are constants. However, the raw retinotopic coordinates $v$ is noisy. It will violate the topological condition by simply enforcing the coordinate $\hat{u}$ to the noisy coordinates. We proposed a method to generate a smooth and topological parametrization. Mathematically, the problem is to find the minimum of energy,

$$(\hat{u}, \hat{v}) = \arg\min E = \int (\hat{\rho} - k_1 \hat{v}^{(1)} - b_1)^2 + (\hat{\theta} - k_2 \hat{v}^{(2)} - b_2)^2$$
$$+ \lambda_1 |\nabla_{\hat{u}} \hat{v}|^2 + \lambda_2 w |\hat{v} - v|^2 ds, \quad s.t. |\mu_{u \to \hat{u}}| < 1, \qquad (1)$$

where $\hat{u} = (\hat{u}^{(1)}, \hat{u}^{(2)})$ is the desired parametrization coordinates, $\hat{v} = (\hat{v}^{(1)}, \hat{v}^{(2)})$ is the desired smoothing retinotopic coordinates, $\nabla_{\hat{u}}$ is the gradient operator defined on the parametric domain $\hat{u}$, i.e. $\nabla_{\hat{u}} = (\partial/\partial \hat{u}^{(1)}, \partial/\partial \hat{u}^{(2)})$, $\lambda_1, \lambda_2$ are constants, $w$ is a weight coefficient with the purpose of emphasis high-quality points, and $\mu_{u \to \hat{u}}$ is the Beltrami coefficient [17] associated with the mapping from the initial parametrization $u$ to the desired $\hat{u}$. The first two terms in Eq. (1) are introduced to linearly align the parametric coordinates to the smoothed retinotopic coordinates. Likewise, the last two terms are introduced with the purpose of smooth the noisy $v$. Lastly, the constraint, $|\mu_{u \to \hat{u}}| < 1$, is introduced to ensure the new parametric coordinates is topological respect to the cortical surface [18].

## 2.3 The Iterative Parametrization.

The optimization of energy in Eq. (3) is difficult as the $\hat{u}$ and $\hat{v}$ have influences on each other. We adopt the ADMM [19] to separate the problem into two sub-problems, namely the smoothing problem and the parametrization problem, then solve the sub-problems iteratively. In practice, the cortical surface is discretized as triangular mesh consists of triangular faces and vertices, denoted by $S = (F, V)$. The retinotopic parameters $(v, \sigma, R)$ are solved by the pRF method [4, 20] for each vertex. We denote the surface with retinotopic parameters by $(F, V, v, R)$, which is the input of the method.

The overall pipeline of the parametrization is illustrated in **Fig. 2**. (1) Given the surface with retinotopic parameters, as illustrated in **Fig. 2**(a), we first find the region of interest (ROI). In this paper, we select the region of interest as the V1/V2/V3 complex; (2) Then we find a geodesic patch that contains the ROI. Specifically, pick a point as the center, find geodesic distances from the center point to all vertices, and keep the portion of the surface whose geodesic distance is within certain value $r$. We call this patch as the geodesic patch, $P = (F_P, V_P, v_P, R_P)$. The purpose to bound the patch by geodesic distance is to reduce distance distortion; (3) Later, we parametrize the patch to unit disk by conformal [21] mapping $c$, denoted by $u = c(P)$. This is the initial coordinates for our method, as illustrated in **Fig. 2** (b); (4) The retinotopic coordinates are smoothed on the domain of $u$. Specifically, with the initial parametrization $u$ and the raw retinotopic coordinates $v$, fitting a smooth function $\hat{v}(u)$ to approach the raw coordinates. **Fig. 2**(d) shows the smoothed result from the raw result **Fig. 2**(c) in the ROI; (5) Adjust the parametric coordinate from $u$ to $\hat{u}$ such that $\hat{u}'s$ polar coordinates linearly relate to the visual coordinates $v$, illustrated in **Fig. 2**(e). During the adjusting, we will ensure the adjusted coordinates still have one-to-one mapping to the cortical surface, namely keeping the topological condition by enforcing $|\mu_{u \to \hat{u}}| < 1$; (6) The adjusting step (5) will break the ideal linearity, so procedure (4)(5) are repeated until the error is within tolerance. The key part of the pipeline is the Error-Tolerant Teichmüller Map (ETTM) in Step (5). The idea to move coordinates from $u$ to the desired location $\hat{u}$ is by setting landmarks and move accordingly.

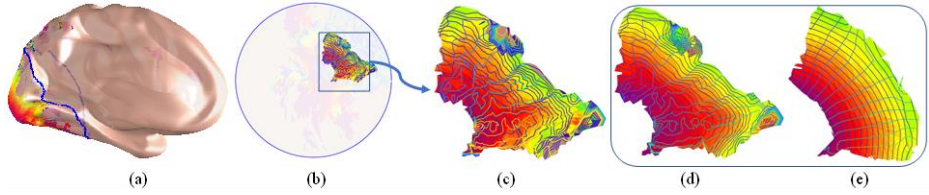

**Fig. 2.** The pipeline of the iterative parametrization: (a) Select the ROI, i.e. V1/V2/V3 complex and compute a geodesic disk patch contains the ROI; (b) Map the patch (enclosed by the blue curve in the occipital lobe) to the 2D unit disk; (c) A zoomed view of the ROI; (d) Apply Laplacian smoothing on the retinotopic data; (e) Adjust the parametric coordinates by ETTM according to the smoothed data. The ellipse which encloses (d)(e) means the steps are repeated.

**Error Tolerant Teichmüller Map.**

To move coordinates toward the target, we set some landmarks. Specifically, all the points within the ROI are selected as landmarks. The landmark target is set by $\hat{u}^{(1)} = k_1 \hat{v}^{(1)} + b_1$ (similar for $u^{(2)}$). Although $\hat{v}$ is smoothed, the landmarks may still have errors. Previous work, e.g. Teichmüller map (T-map) [22] cannot handle this situation. The Error Tolerant Teichmüller Map (ETTM) is proposed to enhance the T-map to tackle landmarks with errors. The key idea of ETTM is to check the topology condition, $|\mu_{u \to \hat{u}}| < 1$, and seek the most similar alternative parametrization that is topological. $\mu_{u \to \hat{u}}$ is defined as $\mu_{u \to \hat{u}} = \left(\frac{\partial \hat{u}}{\partial u^{(1)}} + i \frac{\partial \hat{u}}{\partial u^{(2)}}\right) / \left(\frac{\partial \hat{u}}{\partial u^{(1)}} - i \frac{\partial \hat{u}}{\partial u^{(2)}}\right), i = \sqrt{-1}$. The partial derivatives are approximated by piecewise linear interpretation of the discrete values. If the topology is violated, i.e. $|\mu_{u \to \hat{u}}| > 1$, we will shrink the magnitude, $\mu'_{u \to \hat{u}} = \alpha \mu_{u \to \hat{u}}$ to ensure $|\mu'_{u \to \hat{u}}|$ close but less than 1. $\mu'_{u \to \hat{u}}$ corresponds to the topological map that is closest to the previous non-topological map. Once the proper $\mu'_{u \to \hat{u}}$ is given, we recovery from $\mu'_{u \to \hat{u}}$ to $\hat{u}$ by Linear Beltrami Solver (LBS). We explain the LBS in brief and refer readers to [22] for the details. Denote $\mu'_{u \to \hat{u}} = \rho + i\tau$, according to the definition, we have $-\frac{\partial \hat{u}^{(1)}}{\partial u^{(2)}} = \alpha_1 \frac{\partial \hat{u}^{(2)}}{\partial u^{(1)}} + \alpha_2 \frac{\partial \hat{u}^{(2)}}{\partial u^{(2)}}$ and $\frac{\partial \hat{u}^{(1)}}{\partial u^{(1)}} = \alpha_1 \frac{\partial \hat{u}^{(2)}}{\partial u^{(1)}} + \alpha_2 \frac{\partial \hat{u}^{(2)}}{\partial u^{(2)}}$, where $\alpha_1 = \frac{(\rho-1)^2 + \tau^2}{1 - \rho^2 - \tau^2}$, $\alpha_2 = -\frac{2\tau}{1 - \rho^2 - \tau^2}$ and $\alpha_3 = \frac{1 + 2\rho + \rho^2 + \tau^2}{1 - \rho^2 - \tau^2}$. Apply $\frac{\partial}{\partial u^{(1)}}$ on the first equation, and plus $\frac{\partial}{\partial u^{(2)}}$ on the second one, one can write, $\nabla \cdot A \nabla \hat{u}^{(1)} = 0$, where $A = \begin{pmatrix} \alpha_1 & \alpha_2 \\ \alpha_2 & \alpha_3 \end{pmatrix}$.

Then one can solve $\hat{u}^{(1)}$ with boundary conditions. Similarly, $\hat{u}^{(2)}$ can be solved. We summarize ETTM in **Alg. 1**, and the overall procedure in **Alg. 2**.

**Algorithm 1**. Error Tolerant Teichmüller Map

**Input**: Surface $(F, V)$, initial coordinates $u$, and error-existed landmarks $\{l_i, \hat{T}_i\}$.
**Result**: $\hat{u}$ with uniform distortion and minimal landmark misalignment.
**Initialize**: $\hat{u}_i \leftarrow \hat{T}_i$, if $i$ is a landmark point, else $\hat{u}_i = u_i$.
    **repeat**
        1. Compute $\mu_{u \to \hat{u}}$, and chop $\mu'_{u \to \hat{u}} = \alpha \mu_{u \to \hat{u}}$ for those $|\mu'_{u \to \hat{u}}| < 1$.
        2. Uniform $\mu''_{u \to \hat{u}} = \beta \mu'_{u \to \hat{u}}$, with $\beta = mean(|\mu'_{u \to \hat{u}}|)$.
        3. Recovery $\hat{u}$ the with $\mu''_{u \to \hat{u}}$ and landmarks $\{l_i, \hat{T}_i\}$ by LBS.
    **until** $|\mu_{u \to \hat{u}}| < 1$ and $||\mu_{u \to \hat{u}}| - \beta| < \epsilon$
**return** $\hat{u}$.

**Algorithm 2**. Iterative Parametrization

**Input**: Retinotopic coordinates on the surface $(F, V, v^{(1)}, v^{(2)}, R)$, and radius $r$.
**Result**: Parametrized coordinates $\hat{u}$, that linearly aligns to visual coordinates.
    1. Pick a point as center and compute the geodesic distance to this point.
    2. Keep the portion of surface within $r$, denote as $P = (F_p, V_p, v_p^{(1)}, v_p^{(2)})$
    3. Compute initial $u = c(P)$, and initialize $\hat{u} = u$, and $\hat{v} = v$.
    **repeat**
        4. Smoothing the eccentricity $\hat{v}^{(1)}$ to get new $\hat{v}^{(1)}$.
        5. Determine $k_1, b_1$ and adjust $\hat{\rho}$ to enforce $\hat{\rho} = k_1 \hat{v}^{(1)} + b_1$.
        6. Set landmarks and apply the ETTM to get new $(\hat{u}^{(1)}, \hat{u}^{(2)})$.
        7. Smoothing the polar angle $\hat{v}^{(2)}$ to get new $\hat{v}^{(2)}$.
        8. Determine $k_2, b_2$ and adjust $\hat{\theta}$ to enforce $\hat{\theta} = k_2 \hat{v}^{(2)} + b_2$.
        9. Set landmarks and apply the ETTM to get new $(\hat{u}^{(1)}, \hat{u}^{(2)})$.
    **until** $\max|\hat{\rho} - k_1 \hat{v}^{(1)} - b_1| < \epsilon$ and $\max|\hat{\theta} - k_2 \hat{v}^{(2)} - b_2| < \epsilon$
**return** $\hat{u}, \hat{v}$.

## 3 Dataset and Evaluation Method

### 3.1 Synthetic Data

The problem of the real dataset is there is no ground truth. We generate a synthetic dataset to compare the performance. Specifically, we take the average of cortical surface (with cortical registration) in the Human Connectome Project (HCP) [20] dataset (only use the anatomical surface), map it to the 2D by orthographical projection. Then we generate the ground truth retinotopic coordinates by the complex-log-model with the formula, $u = k \ln\left(\hat{v}^{(1)} e^{i\hat{v}^{(2)}} + a\right)$, with $k = 1, a = 1$. The complex-log-model is a good approximation of the retinotopic mapping introduced by Schwartz [23]. Finally, add noise to $\hat{v}$, i.e. $v = \hat{v} + noise$.

### 3.2 Evaluation Metric

The evaluation is, given noisy $v$, which algorithm can recover best visual coordinates respect to the ground truth $\hat{v}$, cortical magnification, and the compatibility to neuroscientific results. (1) The visual coordinates error is computed with Euclidean distance; (2) The cortical magnification factor $M$ is defined as the area of corresponding cortical surface $A(P)$ divided by the area of visual field $A(\hat{v})$ for a small patch $\sigma$ around the center $\hat{v}$, namely $M = \lim_{\sigma \to 0} \frac{A(P)}{A(\hat{v})}$. In the discrete triangular mesh $(F, V)$, for each point $P \in V$, we select the dual cell for $P$ as the patch $\sigma$. Then compute the dual cell area $A(P)$, and its corresponding area in visual field $A(\hat{v})$. (3) We use the number of non-topological points $\tilde{T}$, i.e. the point moves out of the polygon consisted of neighbors' points, to quantify the compatibility. The ideal result is no such non-topological points exist, i.e. $\tilde{T} = 0$.

### 3.3 HCP Dataset

The Human Connectome Project (HCP) is a high-quality retinotopy dataset [20]. The data is collected on a modern MRI machine of 7 Tesla magnetitic field with carefully designed visual stimuli. Even so, the result is still noisy. We also applied our method and compare it with harmonic parametrization [24], orthographic projection [12], angle-preserving parametrization [21], area-preserving parametrization [25], and nearly isometric parametrization [26], based on same smoothing setting.

## 4 Result

### 4.1 Synthetic data

We first show the method on synthetic data. The proposed method is compared with other parametrization methods with the same smoothing setting. Methods and their performances are listed in **Tab. 1**. We find: (1) the proposed method can recover the best visual coordinates, evaluate the best cortical magnification factor, and most compatible with neuroscientific results; (2) Angle-preserving and the nearly isometric methods are in the second rank. However, the nearly isometric method is more time consuming; (3) Area-preserving map, harmonic map, and orthographic projection are in the third rank with similar performance.

| Method | $\hat{v}$ error | | $M$ error | | $\tilde{T}$ | Time/s |
|---|---|---|---|---|---|---|
| | Avg. | SD | Avg. | SD | | |
| Raw Data | 0.385/0.694 | 0.225/0.392 | 27179/27628 | 13602/13788 | 645/680 | **0.0/0.0** |
| Harmonic | 0.343/0.616 | 0.205/0.362 | 26992/27567 | 13518/13759 | 646/685 | 1.9/1.8 |
| Orthographic | 0.346/0.620 | 0.201/0.355 | 27022/27580 | 13550/13770 | 645/686 | 2.0/2.0 |
| Angle-Preserving | 0.341/0.614 | 0.205/0.359 | 26985/27567 | 13514/13761 | 650/689 | 2.3/2.2 |
| Area-Preserving | 0.342/0.618 | 0.198/0.351 | 27019/27576 | 13560/13776 | 649/683 | 16.2/16.2 |
| Nearly-Isometric | 0.339/0.613 | 0.197/0.353 | 27010/27576 | 13567/13774 | 649/683 | 32.1/33.5 |
| **Proposed** | **0.142/0.216** | **0.119/0.136** | **18728/21422** | **12291/12629** | **564/677** | 7.1/6.4 |

**Table 1.** Compare different methods by three metrics, visual coordinate $\hat{v}$ difference, the cortical magnification factor $M$ difference relative to the ground truth, and the number of topology violations. The metrics are evaluated for a small noise level (PSNR=20) and a big noise level (PSNR=10) (with the separation symbol '/'). "Avg." is the average difference, and "SD" is the standard deviation of the difference.

### 4.2 HCP data

We apply the proposed algorithm to the HCP data. The raw retinotopic maps of the first subject (in the left hemisphere) is shown in **Fig. 3**(a). **Fig. 3**(b)-(c) are for angle-preserving and area-preserving respectively. The proposed method is in **Fig. 3**(d). The same data are overlaid on the inflated cortical surface in **Fig. 3**(e)-(h), respectively. Visually, the proposed result is smoother and closer to neuroscientific results.

Although no ground-truth is available, we try to evaluate the results indirectly by two aspects: (1) whether the smoothed result is compatible with the neuropsychology result; and (2) whether the cortical magnification factor (CMF), agrees to public records. For the first subject, we report the violation numbers are: $\tilde{T} = 189$ (raw), $\tilde{T} = 90$ (harmonic), $\tilde{T} = 101$ (orthographic), $\tilde{T} = 85$ (angle-preserving), $\tilde{T} = 120$ (area-preserving), $\tilde{T} = 83$ (nearly-isometric), and $\tilde{T} = \mathbf{20}$ (proposed). We see the method generated the least number of topology violations, which means the proposed method generates more reasonable results than other methods.

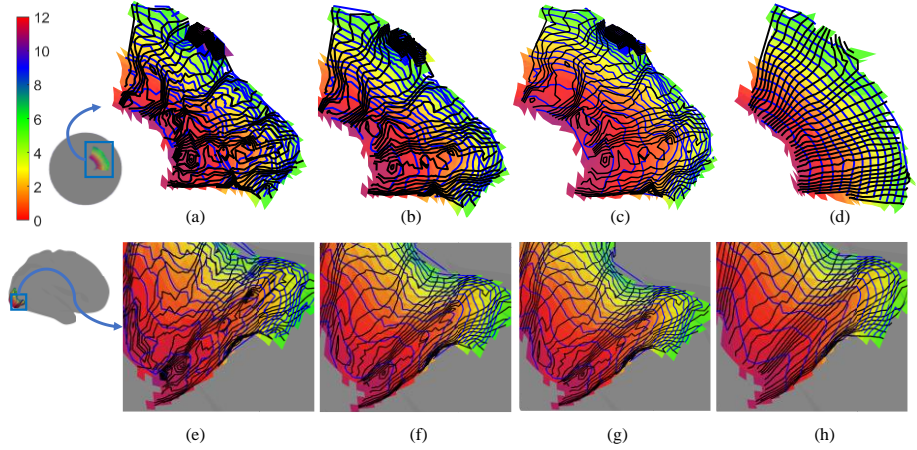

**Fig. 3.** First subject's visual coordinates contour: (a) The conformal parameterization with the raw retinotopic coordinates; (c) Smoothed on the conformal parameterization; (d) Smoothed based on area-preserving parametrization; (d) Our method; (e)-(h) Results on the inflated surface by same data of (a)-(h)'s, respectively. Adjacent blue lines are drawn with an 0.5° eccentricity interval, and the black lines are drawn with a 10° polar angle interval.

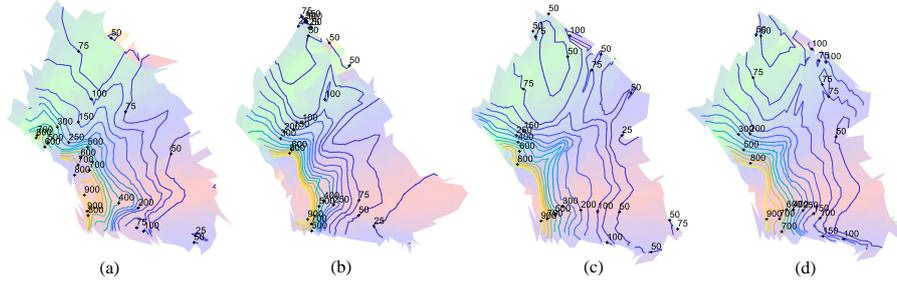

**Fig. 4.** (a)-(c): the first three CMF; (d) The average CMF for the first three subjects.

We further estimate CMF by our method. **Fig. 4**(a)-(c) shows the first three CMF overlaid on the parametric space. The extra benefit of the proposed method is the parametric coordinates have been aligned if the constants $k, b$ are the same for all subjects. So, our method can directly take an average of the parametric domain across the subjects, which is shown in **Fig. 4**(d). Previous methods [12, 27] can only estimate CMF as a function of eccentricity in the periphery. We provide the first quantification of CMF along with different polar angles near the fovea. Also, the results are consistent with the overall observation that the visual field is compressed vertically [28].